\documentclass[iop, twocolumn]{aastex6}
\hypersetup{citecolor=blue,urlcolor=red}
\usepackage{amsmath}
\usepackage{url}
\usepackage{enumerate}
\usepackage{multirow}

\shorttitle{Single pulse study of PSR J0621+1002}
\shortauthors{Wang et al.}

\begin{document}
\title{A single pulse study of a millisecond pulsar PSR J0621+1002}
\author
{S. Q. Wang\altaffilmark{1,2,3}, J. B. Wang\altaffilmark{1,4,5}, N. Wang\altaffilmark{1,4,5}, Y. Feng\altaffilmark{6,2}, S. B. Zhang\altaffilmark{7}, K.J. Lee\altaffilmark{8}, D. Li\altaffilmark{6,2,9}, J. G. Lu\altaffilmark{6}, J. T. Xie\altaffilmark{1,2}, D. J. Zhou\altaffilmark{6,2}, L. Zhang\altaffilmark{6}}

\altaffiltext{1}{Xinjiang Astronomical Observatory, Chinese Academy of Sciences, 150 Science 1-Street, Urumqi 830011, People's Republic of China;}
\altaffiltext{2}{University of Chinese Academy of Sciences, Beijing 100049, People's Republic of China}
\altaffiltext{3}{CAS Key Laboratory of FAST, NAOC, Chinese Academy of Sciences, Beijing 100101, People's Republic of China}
\altaffiltext{4}{Key Laboratory of Radio Astronomy, Chinese Academy of Sciences, 150 Science 1-Street, Urumqi 830011, People's Republic of China}
\altaffiltext{5}{Xinjiang Key Laboratory of Radio Astrophysics, 150 Science1-Street, Urumqi, Xinjiang, 830011, People's Republic of China}
\altaffiltext{6}{National Astronomical Observatories, Chinese Academy of Sciences, A20 Datun Road, Chaoyang District, Beijing 100101, People's Republic of China}
\altaffiltext{7}{Purple Mountain Observatory, Chinese Academy of Sciences, Nanjing 210008, People's Republic of China}
\altaffiltext{8}{Kavli Institute for Astronomy and Astrophysics, Peking University, Beijing 100871, People's Republic of China}
\altaffiltext{9}{NAOC-UKZN Computational Astrophysics Centre, University of KwaZulu-Natal, Durban 4000, South Africa}

\email{wangjingbo@xao.ac.cn, na.wang@xao.ac.cn}

\begin{abstract}

We present radio observation of a millisecond pulsar PSR J0621+1002 using the Five-hundred-meter Aperture Spherical radio Telescope (FAST). The pulsar shows periodic pulse intensity modulations for both the first and the third pulse components. The fluctuation spectrum  of the first pulse component has one peak of 3.0$\pm$0.1 pulse periods, while that of the third pulse component has two diffused peaks of 3.0$\pm$0.1 and 200$\pm$1 pulse periods. The single pulse timing analysis is carried out for this pulsar and the single pulses can be divided into two classes based on the post-fit timing residuals. We examined the achievable timing precision using only the pulses in one class or bright pulses. However, the timing precision improvement is not achievable.

\end{abstract}


\section{INDRUCTION}



Radio pulsars are known to exhibit various emission properties, such as nulling \citep{Backer70a, Wang07}, mode changing~\citep{Bartel82} and giant pulse~\citep{Staelin1968}. 
Some pulsars exhibit periodic emission variations~\citep{Weltevrede2016}. 
One well known of such phenomenon is subpulse drifting, in which the subpulse drifts in pulse longitude across a pulse sequence~\citep{Weltevrede2016}. 
Another similar phenomenon is amplitude  modulation, where pulses only show periodic intensity modulations that do not propagate in pulse phase~\citep{Basu16}.

Periodic amplitude modulation and drifting subpulse are mainly seen in normal pulsars. 
These phenomena are difficult to detect in millisecond pulsars (MSPs) because of their low flux densities~\citep{Edwards03}. 
Single-pulse studies have been only carried out for some bright MSPs~\citep{Jenet98, Jenet01,Liu15, Wang20}. 
\citet{Edwards03} reported pulse-to-pulse intensity modulations for six MSPs. 
\citet{Liu16} revealed the diffused subpulse drifting phenomenon of PSR J1713+0747. 
Recently, \citet{Mahajan18} found the first mode changing MSP, PSR B1957+20.

It is possible to detect nanohertz gravitational waves by monitoring pulse times of arrival (ToAs)  of an ensemble of the
most stable MSPs~\citep{Shannon13, Arzoumanian20}. 
The success of this experiment strongly depends on the achievable timing precision. On short time scales, pulsar timing precision is limited by white noise~\citep{Liu12, Shannon14,Lam19}.  Radiometer noise and phase variation of integrated profiles induced by pulse-to-pulse variability, commonly referred to as jitter noise, are the main sources of white noise. Thus, single-pulse studies of MSPs can provide us a fundamental limit on the achievable timing precision on short timescales.
Highly sensitive radio telescopes, such as the Five hundred meter Aperture Spherical Telescope (FAST), provide us a great opportunity to study pulse-to-pulse variability of MSPs and jitter noise.

PSR J0621+1002 is a 28.9\,ms MSP monitored by the European Pulsar Timing Array (EPTA, see~\citealt{Desvignes16}).
The pulsar is in a binary system with an 8.3-days orbital period and a 0.41\,$\rm M_{\odot}$ CO white dwarf companion~\citep{Camilo96, Kramer98, Splaver02, Kasian12}.
In this paper, we present analysis on periodic pulse intensity modulation and jitter noise of the pulsar with the FAST.
In Section 2, we describe our observation.  The results are presented in section 3. We discuss and summarise our results in Section 4.

\section{OBSERVATIONS AND DATA PROCESSING}


We obtained 54066 single pulses in a 26 minutes observation on  January 14 2020 (project ID 3062).
The observation was conducted using the central beam of the 19-beam receiver with a frequency range between 1050 and 1450 MHz \citep{Jiang19}. 
The data were recorded in search mode PSRFITS format  with 4 polarizations, 8-bit samples of an 49.152-$\mu$s interval, and 4096 frequency channels.

Individual pulses were extracted using {\sc dspsr} software package (\citealt{Straten11}). The $-$K option  in {\sc dspsr} was used to remove  inter-channel dispersion delays. 5\% of the band-edges and flag narrow-band and impulsive radio-frequency interference (RFI) are removed using {\sc psrchive} software package (\citealt{Hotan04}). More details of the RFI environment of FAST are provided in~\citet{Jiang19}.
Polarization calibration was achieved by correcting for the differential gain and
phase between the receptors through separate measurements using a noise diode signal.
Flux density  was calibrated using observations of 3C 286~\citep{Baars77}.
Flux density and rotation measure (RM) are obtained using the {\sc psrchive} program {\sc psrflux} and {\sc rmfit}. 

Noise-free standard templates were formed by interactively fitting scaled von Mises functions
(using the {\sc psrchive} program {\sc paas}) to a high S/N observed profile. 
ToAs  were obtained by cross-correlating pulse profile with the template using the {\sc pat} command.  Timing residuals were calculated using the {\sc tempo2} software package ~\citep{Hobbs06}. 
PSRSALSA package was used to carry out fluctuation  analysis ~\citep{Weltevrede2016}.

\section{RESULTS}

\subsection{The periodical pulse intensity modulation }

\begin{figure}
  \centering
  \includegraphics[width=80mm]{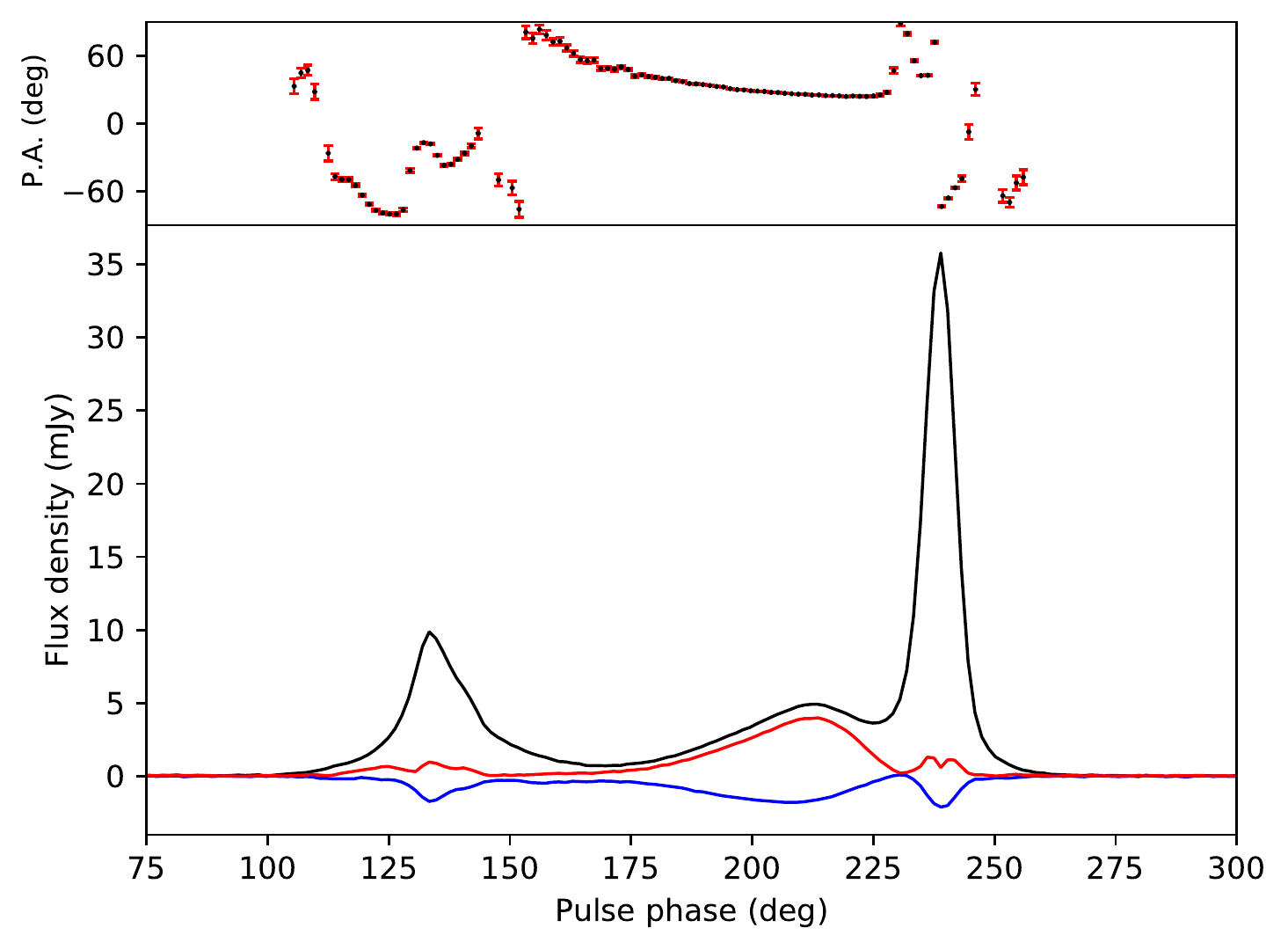}
  \caption{Polarization profile for PSR J0621+1002. The black, red, and blue lines are the total intensity, linear polarized intensity, and circular polarized intensity, respectively.  Position angles (black dots) and corresponding uncertainties (red bars) of the linear polarized emission are shown as a function of pulse phase. }
 \label{pola}
\end{figure}

\begin{figure}
  \centering
  \includegraphics[width=80mm]{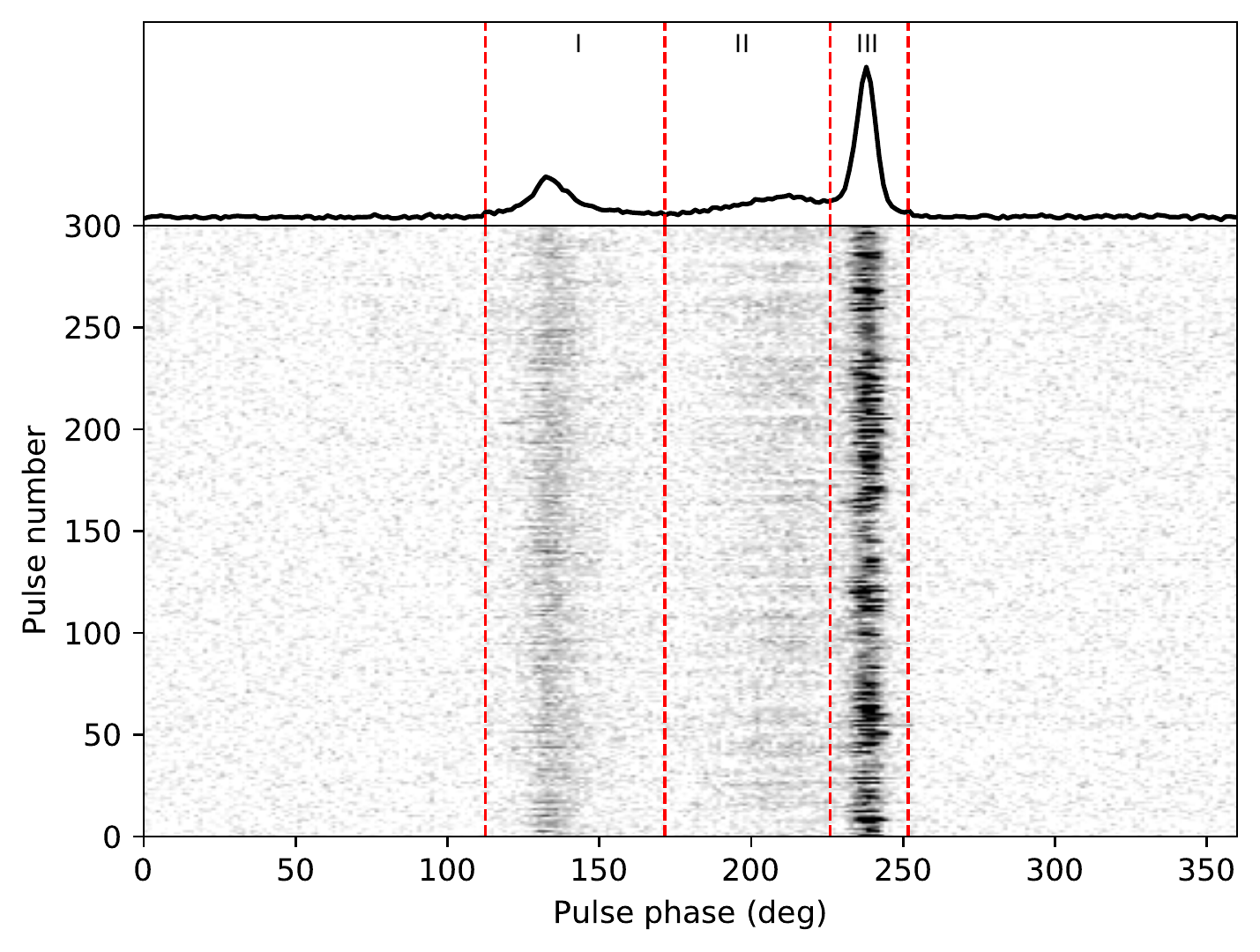}
  \caption{Single-pulse stack of 300 pulses of PSR J0621+1002.  Average pulse profile is shown in upper panel and divided into three components (labeled as I, II and III) by the four vertical lines. 
  }
 \label{gray}
\end{figure}

\begin{figure*}
  \centering
  \includegraphics[width=80mm]{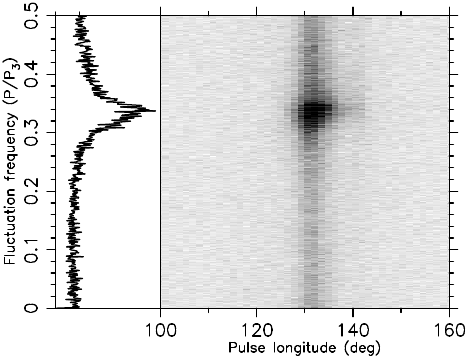}
  \includegraphics[width=80mm]{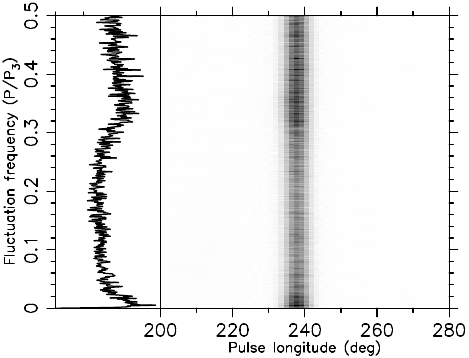}
  \includegraphics[width=80mm]{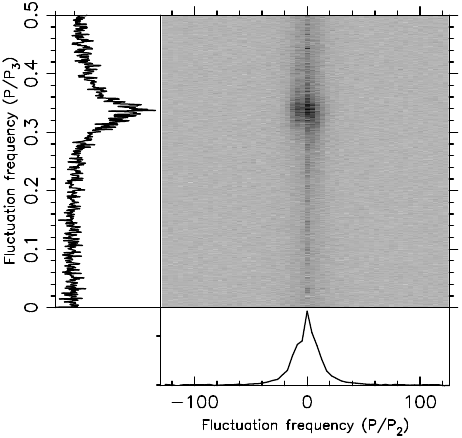}
  \includegraphics[width=80mm]{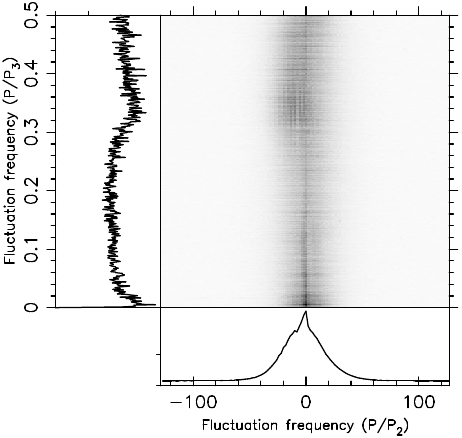}
  \caption{Fluctuation analysis of the emission of PSR J0621+1002. Top row: the LRFS and a side panel showing the horizontally integrated power for the first (left panel) and third (right panel) pulse components.Bottom row: the 2DFS and a side panel showing the horizontally integrated power for the first (left panel) and third (right panel) pulse components.}
  \label{drift}
\end{figure*}

The polarization profile for PSR J0621+1002 is shown in Figure~\ref{pola}. The overall pulse width is about 175 degrees, nearly half of the pulse period. The profile features are consistent with the EPTA observation at  1.4 GHz ~\citep{Desvignes16}. The pulse profile has three components and the fractional linear polarization for the second component is higher than the first and third components.

The position angles (PAs) vary significantly across the first and the third components and it is relatively smooth across the second component. 
Generally, for pulse profile with three components, the first and third pulse components are related to the cone emissions and the second pulse component is associated with the core emission~\citep{Backer76, Rankin83} and the gradient of PA swing is large near the pulse centre~\citep{Lyne88}.
The second component of PSR J0621+1002 exhibits a relatively smooth PA swing which does not agree with the core emission. This component also shows high fractional linear polarization which is characterized by the cone emission. Therefore, we suggested that all the three components of PSR J0621+1002 are related to the cone emissions.
Our measured flux density at 1250 MHz and RM  are  1.90\,mJy and $53.0\pm0.1\,{\rm rad \,m^{-2}}$, respectively, which are consistent with the previously published results~\citep{Sobey2019}.

Single-pulse stack of 300 pulses is shown in Figure~\ref{gray}.
To investigate the pulse modulation behavior,  longitude-resolved fluctuation spectrum (LRFS, \citealt{Backer70b}) and two-dimensional fluctuation spectrum (2DFS, \citealt{Edwards02}) was carried out for each pulse component.
2DFS is a useful tool to determine whether subpulses are
drifting in pulse longitude~\citep{Weltevrede2016} and LRFS can be used to detect periodicity of subpulse modulation.
Using {\sc PSRSALSA} package with FFT-size of 1024, the LRFS and 2DFS of all the data are shown in the upper and bottom panels of Figure~\ref{drift}, respectively.
There is a clear periodic pulse intensity modulation for both the first and the third pulse components.
The fluctuation spectrum of the first pulse component has a clear peak of $3.0\pm0.1\,P$ with the pulse period $P$, which means that the vertical separation of drift bands of this component is coherent with $P_3=3.0\pm0.1 \,P$.
However, the fluctuation spectrum of the third pulse component is diffused at the peaks of $3.0\pm0.1\,P$ and 200$\pm$1\,$P$ which suggests that the vertical separation of drift bands is not fully coherent.
The asymmetric feature (along X=0) in the 2DFS of the first component suggests the existence of pulse drifting across the pulse longitude. However, our resolution is not enough to fully resolve the feature and yield a measurement of $P_2$.

The $P_3$ for the first and  third components seem identical. Correlation analysis between them was carried out to check whether the intensity variation following this period is phase-locked (more details see \citealt{Kou21}). We found no evidence that the intensity variations for the first and third pulse components are phase-locked.

\subsection{Pulse energy distribution}

\begin{figure}
  \centering
  \includegraphics[width=80mm]{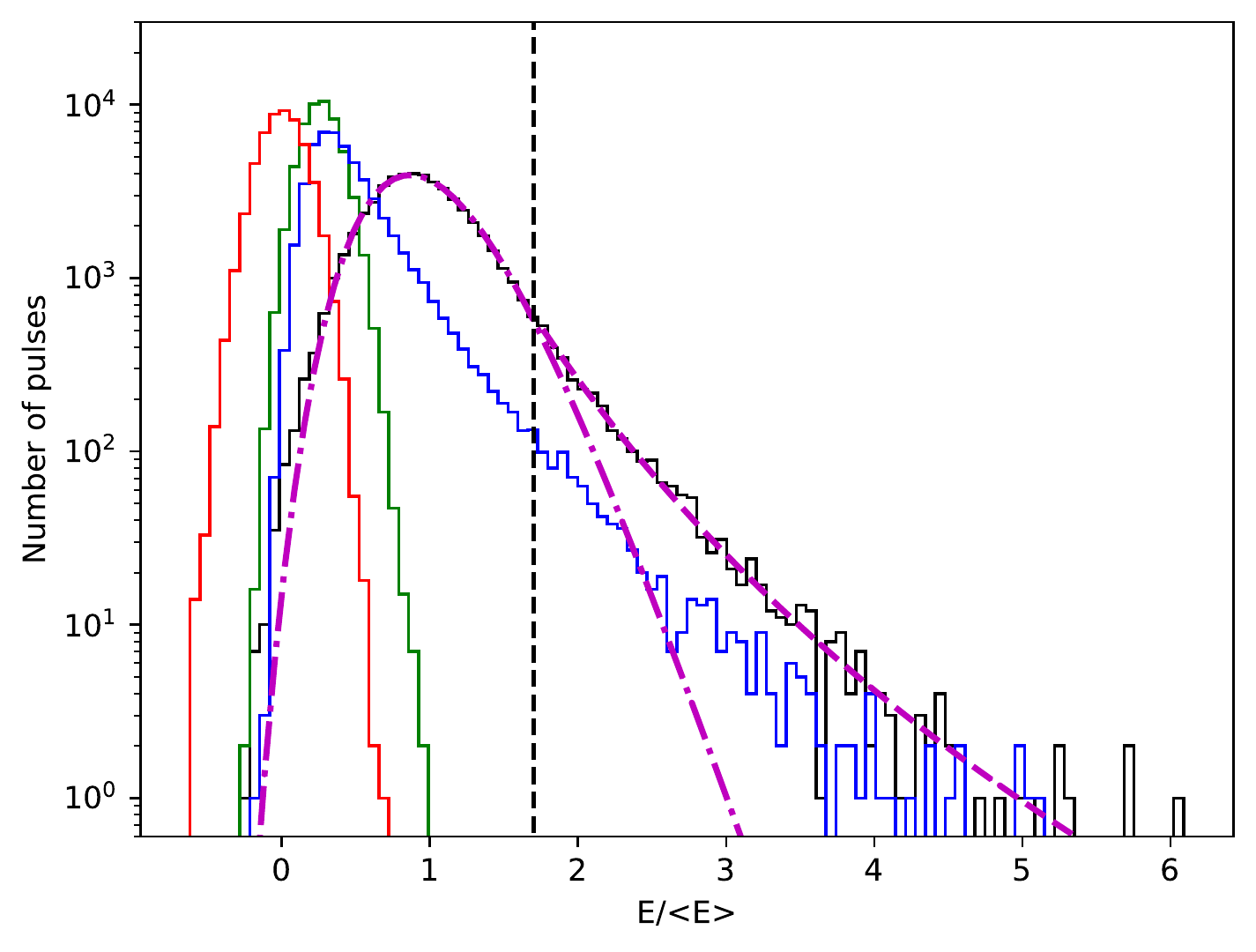}
  \caption{Normalised pulse energy distribution for PSR  J0621+1002. The red, black, green and blue lines are the pulse energies of off$-$pulse and on$-$pulse region for the first and the third pulse components, respectively. The dashed-dotted and dashed lines are the log-normal and power-law fitting functions of the pulse energies for on-pulse region, respectively. The vertical black line is the energy cutoff at 1.7 $E/\langle E \rangle$. Note that the energy distribution is not deconvolved with the noise distribution.}
  \label{energy}
\end{figure}

Pulse energy distributions of the off-pulse and on-pulse regions for the first  and the third pulse components are presented in Figure~\ref{energy}.
All the energies are normalized by the mean on-pulse energy.
On-pulse region is determined as the longitude range where the pulse intensity significantly exceeds (larger than 3 $\sigma$) the baseline noise and the off-pulse energy is measured from a region with the same duration as the on-pulse region in the baseline.

Pulse energy of the off-pulse region follows a narrow normal distribution, while that of the on-pulse region can be described as log-normal distribution with a high-energy power-law tail. The energy cutoff is at 1.7E/$\langle E \rangle$ for the power-law tail (the vertical black line in Figure~\ref{energy}) and the spectral index is $-8.0\pm0.1$.
Pulse energy distribution of the first pulse component can be fitted well by a log-normal function. Pulse energy of the third pulse component also follows a log-normal distribution with a power-law tail.
Generally, pulse energy for MSPs can be well modeled using log-normal or Gaussian distribution~\citep{Shannon14}.
However, log-normal energy distribution with an excess of high-energy pulses has been detected in many normal pulsars \citep{Mickaliger18}.

\subsection{The jitter noise}

\begin{figure}
  \centering
  \includegraphics[width=85mm]{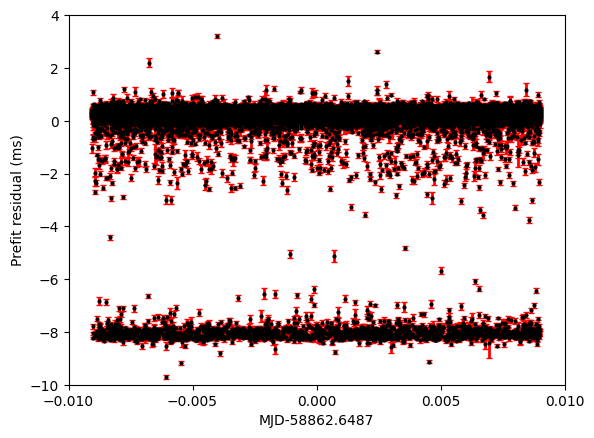}
  \caption{Single pulse timing residuals (black dots) for PSR J0621+1002. The red bars are the ToA uncertainties.
  }\label{res}
\end{figure}

\begin{figure}
  \centering
  \includegraphics[width=85mm]{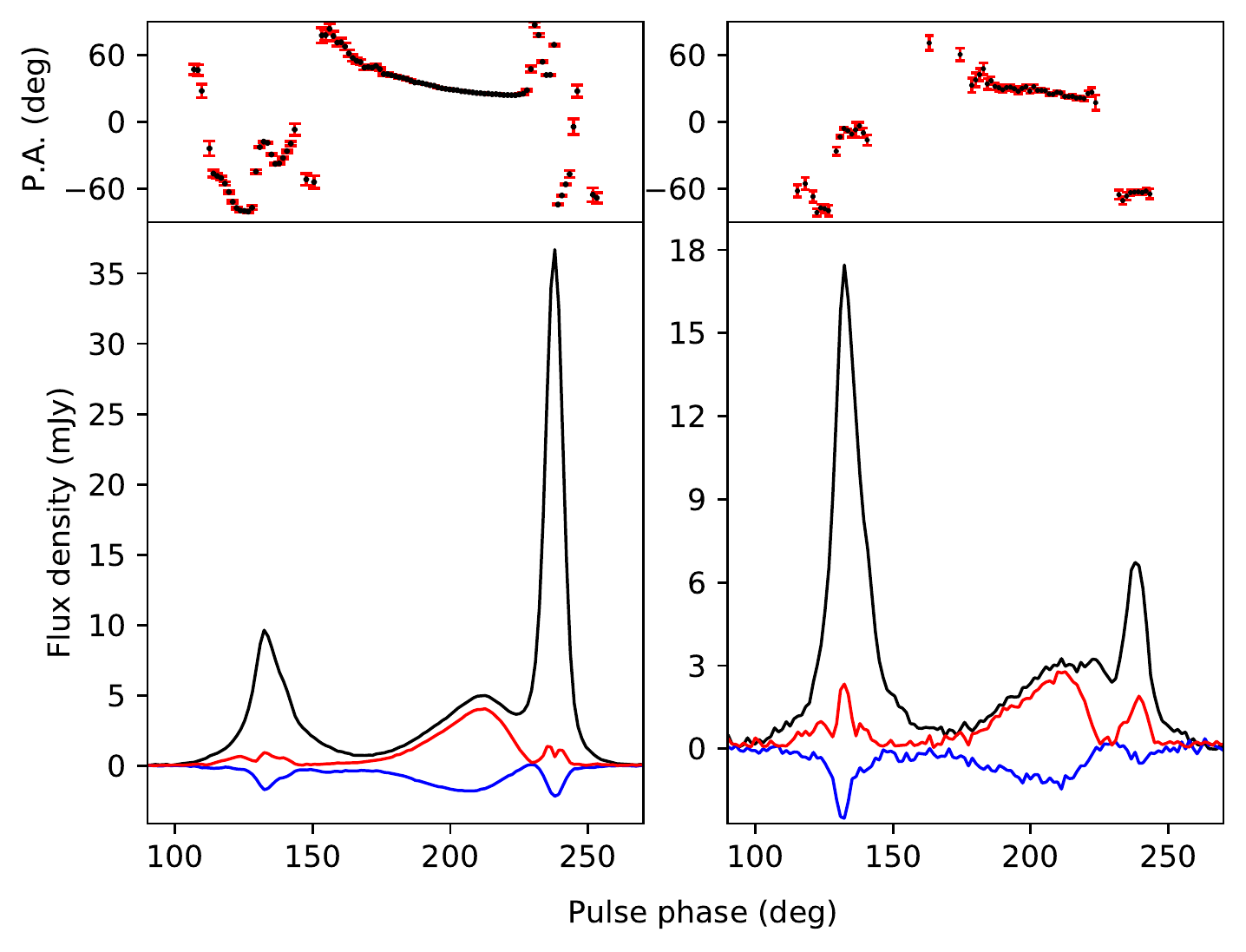}
  \caption{Polarization profiles for the mode A (left panel) and mode B (right panel) of PSR J0621+1002. The black, red, and blue lines are the total intensity, linear polarized intensity, and circular polarized intensity, respectively. Position angle (black dots) and corresponding uncertainties (red bars)  are shown as a function of pulse phase. }
 \label{prof}
\end{figure}

\begin{figure}
  \centering
  \includegraphics[width=85mm]{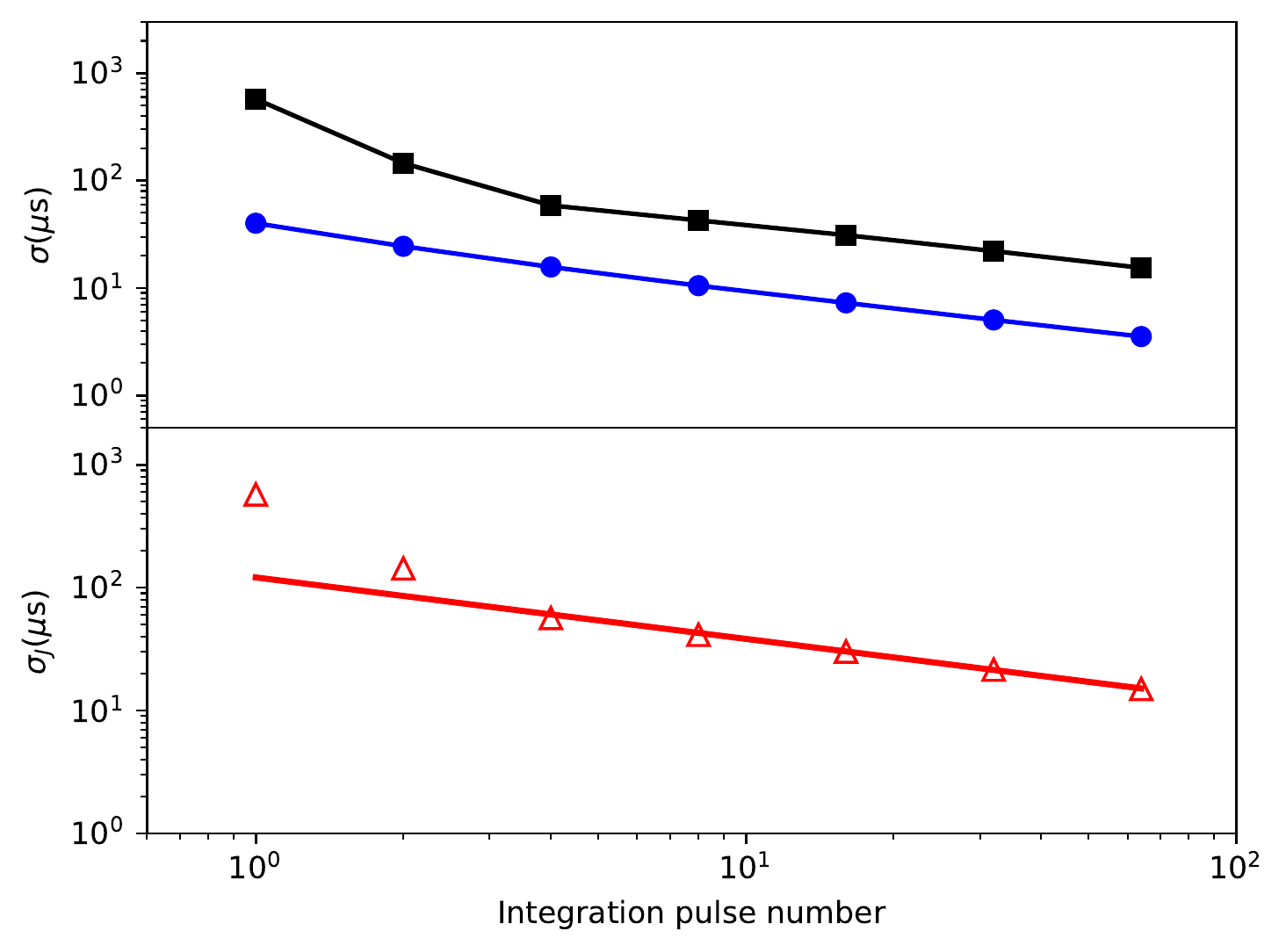}
  \caption{Estimates of jitter noise in PSR~J0621+1002. Upper panel: variations of rms timing residuals  (squares) and ToA uncertainties (circles) versus the number of pulses averaged.
Bottom panel: quadrature difference (triangles) between the rms timing residuals and ToA uncertainties.
The red solid line is the best fitting model for the jitter noise with index of $-0.50\pm0.01$. }\label{jitter1}
\end{figure}

\begin{figure}
  \centering
  \includegraphics[width=85mm]{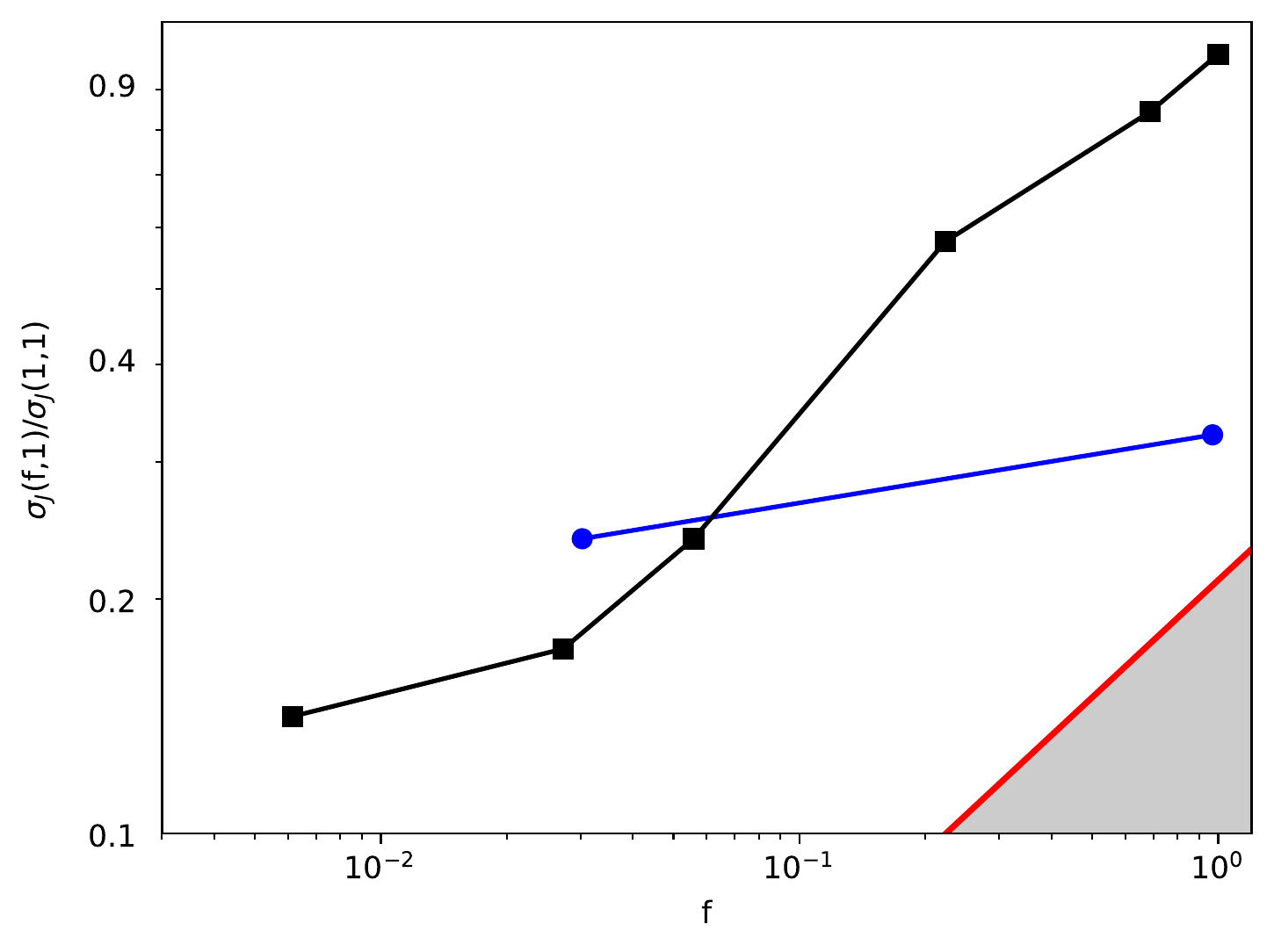}
  \caption{ Fraction of pulses used ($f$) versus normalised pulse jitter ($\sigma_{\rm J}(f,1)/\sigma_{\rm J}(1,1)$) for PSR~J0621+1002. 
  Circles from right to left are the normalized pulse jitters of the pulses in class A and class B, respectively.
  Squares from right to left are the normalized pulse jitters of all the single pulses, single pulses with S/N less than 10, in the range of 10 to 20, 20 to 30, 30 to 50, and larger than 50, respectively. 
  The red line is for the best fitting model for the jitter noise (see the red solid line in Figure~\ref{jitter1}).
  Filled grey area identifies the region $\sigma_{\rm J}(f,1)< \sigma_{\rm J}(1,1)/\sqrt{f}$, in which the improvement on timing precision is achievable using sub-set of pulses with different S/N. }\label{fj}
\end{figure}

ToA uncertainties on short time-scales can be described as~\citep{Cordes10,Liu12,Lam16}: 
\begin{equation}\label{eq}
\sigma^2_{\rm total}=\sigma^2_{\rm rm}+\sigma^2_{\rm J}+\sigma^2_{\rm scint}+\sigma^2_{\rm 0},
\end{equation}
where $\sigma_{\rm rm}$, $\sigma_{\rm J}$, $\sigma_{\rm scint}$ and $\sigma_{\rm 0}$ are the uncertainties induced by radiometer noise, jitter noise,  instability of short-term diffractive scintillation and all other possible contributions. 
$\sigma_{\rm rm}$/$\sigma_{\rm J}$ is proportion to the signal to noise ratio (S/N) of single pulse.
For highly sensitive radio telescopes, such as FAST~\citep{Liu11, Liu12,Hobbs19}, $\sigma_{\rm rm}$ is  expected to be less dominant especially for bright pulsars.

ToA uncertainties induced by short-term diffractive scintillation can be described as (see~\citet{Cordes10}): $\sigma_{\rm scint}^2=\tau ^2/N_{\rm scint}$ with the pulse-broadening time-scale $\tau$ and the number of scintles  $N_{\rm scint}$ in the observation.
The number of scintles  $N_{\rm scint}=(1+\eta \Delta \nu/\nu_{\rm d})(1+\eta \Delta T/ t_{\rm d})$, where $\Delta \nu$ and $\Delta T$ are the observing bandwidth and time, $\nu_{\rm d}$ and $t_{\rm d}$ are the diffractive scintillation bandwidth and time, and $\eta \approx 0.3$ is the scintillation filling factor~\citep{Cordes10}. 
For PSR J0621+1002, the diffractive scintillation bandwidth,  diffractive scintillation time and pulse-broadening time are 1.19\,MHz, 323.4\,s and 0.16\,$\mu$s~\citep{Cordes02}, respectively. In our observation, the observing bandwidth and time are 400\,MHz and 1560\,s, respectively; the $\sigma_{\rm scint}$ is about 10\,ns which is negligible.

Intrinsic single pulse shape and phase variations introduce jitter noise.
We carry out single pulse timing analysis for PSR J0621+1002 and the ephemeris is provided by~\citet{Desvignes16}.
{\sc psrchive} program {\sc paas} was used to fit the integrated profile of the entire observation and formed a noise-free template. 
The ToAs are obtained by cross-correlating single pulse profiles with the standard template. 
The  timing residuals of single pulses are shown in Figure \ref{res} and the post-fit timing residuals are divided into two classes. Single pulses with timing residuals in the range of -5$-$4\,ms and -10$-$-5\,ms are defined as class A and class B, respectively.
The number of pulses in class A and class B are 52429 and 1637, respectively.

The average pulse profiles of class A and class B are shown in Figure~\ref{prof}. It is expected that the average profiles of the two classes are different. The pulse profile of class A has a stronger third pulse component, and that of class B has a stronger first pulse component.
PAs of both classes show complicated variations.
However, there is no clear evidence that the PA swings of  two classes of pulses are different because of the limited signal to noise ratio (S/N).

As mentioned above, $\sigma_{\rm scint}$ in Equation~\ref{eq} is negligible. 
We can assume that all of the excess error in the arrival time measurements are attributed to jitter noise.
Jitter noise can then be obtained by calculating the quadrature difference between the observed rms timing residual and the ToA uncertainty~\citep{Shannon14}: $\sigma_{\rm J}^2{(N_{\rm p})}=\sigma_{\rm obs}^2{(N_{\rm p})}-\sigma_{\rm ToA}^2{(N_{\rm p})}$ where ${N_{\rm p}}$ is the number of averaged pulses. 
Estimates of jitter noise of PSR J0621+1002 are shown in Figure~\ref{jitter1}.
The red solid line in Figure~\ref{jitter1} is the fitted result for the jitter noise scaling $\sigma_{\rm J}{\rm(N_p)} \propto N_{\rm p}^{-1/2}$.
The expected jitter noise for one-hour-long observation is 0.51\,$\mu$s.

The expected timing precision for a fraction $f$ of $N$ pulses: $\sigma_{\rm J}(f,N)=\sigma_{\rm J}(f,1)/\sqrt{fN}$~\citep{Shannon14}, where $\sigma_{\rm J}(f,N)$ is the jitter noise in the $fN$ selected single pulses. 
While using a fraction of pulses, the timing precision improvement is achieved if and only if $\sigma_{\rm J}(f,1)<\sigma_{\rm J}(1,1)\sqrt{f}$.

We examined the achievable timing precision using only the pulses in class A or class B. The pulses in class A and class B and all the single pulses formed three noise-free templates, respectively. ToAs for each class are obtained by cross-correlating the single pulse profile with the corresponding standard template.
As shown in Figure~\ref{fj}, the $\sigma_{\rm J}$ does decrease if only the pulses in one class are selected, but it does not decrease to where the timing precision improvement can be achieved (the filled grey area in Figure~\ref{fj}).

We also examined the achievable timing precision  for PSR J0621+1002 using only bright pulses. All the single pulses are divided into five classes with the S/N less than 10, in the range of 10 to 20, 20 to 30, 30 to 50, and larger than 50.
Note that the S/N is determined by dividing the peak flux density of a single pulse by the rms of the off pulse region.
The pulse numbers of these five classes are 37185, 12050, 3022, 1476, and 333, respectively.
We summed all the single pulses in each class and formed five noise-free templates, respectively. ToAs for each pulse class are obtained by cross-correlating the single pulse profile with the corresponding standard template.
As shown in Figure~\ref{fj},
The $\sigma_{\rm J}$ does decrease for brighter pulses since they originate from a narrower region of the pulse phase.
However, timing precision improvement is not achieved.

\section{ DISCUSSION AND CONCLUSIONS}

PSR J0621+1002 shows periodic pulse intensity modulation for the first and third pulse components. 
The horizontally collapsed fluctuation spectrum of the first pulse component is peaked at 3.0$\pm$0.1 pulse periods, while that of the third pulse component has two diffused peaks of 3.0$\pm$0.1 and 200$\pm$1 pulse periods. No modulation across the pulse phase was found. 
The periodic emission variation has been detected in both normal pulsars and MSPs~\citep{Edwards03,Basu16,Liu16}. 
By analyzing 70 normal pulsars with periodic emission variations, \citet{Basu20} suggested that the physical origin of the periodic amplitude modulation is different from that of the subpulse drifting. Generally, the subpulse drifting phenomenon is only seen in the cone components of the pulse profile, while the periodic amplitude modulation phenomenon is seen in the cone, even both the core and cone profile components~\citep{Basu19}. 
The first and third components of PSR J0621+1002 might associated with the cone emissions, which meets the pattern of \citet{Basu20}.

Two classes of single pulses for PSR J0621+1002 have been distinguished and the average pulse profiles for them are different.
The duration of mode B is only one or two pulse periods (about tens of milliseconds), while that of mode change is typical in the range of several seconds to hours, even much longer~\citep{Wang07, Mahajan18}.
The phenomenon seen in PSR J0621+1002 is pulse jitter that due to radiation instability of single pulse, while mode changing results from the change of pulsar magnetosphere geometries or/and currents~\citep{Timokhin10}.

Jitter noise is currently thought to be a limiting noise process for sensitive radio telescopes~\citep{ Shannon14, Hobbs19}. The jitter noise level for PSR J0621+1002 is 0.51\,$\mu$s for an hour-long observation.
We studied the achievable timing precision  for this pulsar using the pules in class A or class B, and the bright pulses and found that the timing precision improvement is not achievable. 

Similiar, by analyzing the single pulses of PSR J1713+0747, \citet{Liu16} found that no improvement of timing precision is achieved by selecting a sub-set of pulses with specific flux density or pulse width. 
\citet{McKee19} studied the timing properties of giant pulses of PSR B1937+21 and found that although the ToA uncertainties formed by giant pulses are much lower than that of the average pulse profiles, the timing precision is not significantly improved. 
However, the single pulses could still be used to mitigate jitter noise for pulsars whose pulses with different properties or distributions in phase. 
\citet{Kerr15} proposed a new method  which can be used to reduce the jitter noise associated with pulse-to-pulse variability.

\section*{Acknowledgments}
This work made use of the data from the Five-hundred-meter Aperture Spherical radio Telescope, which is a Chinese national mega-science facility, operated by National Astronomical Observatories, Chinese Academy of Sciences.
The FAST telescope is partly supported by The Operation, Maintenance and Upgrading Fund for Astronomical Telescopes and Facility Instruments, budgeted from the Ministry of Finance of China (MOF) and administrated by the Chinese Academy of Science (CAS), and the Key Lab of FAST, National Astronomical Observatories, Chinese Academy of Sciences. 
This work is supported by the National Natural Science Foundation of China (No.12041304), the National SKA Program of China (No.2020SKA0120100), the National Key Research and Development Program of China (No.2017YFA0402600), the Youth Innovation Promotion Association of Chinese Academy of Sciences and the 201* Project of Xinjiang Uygur Autonomous Region of China for Flexibly Fetching in Upscale Talents. 

\software{DSPSR \citep{Straten11}, PSRCHIVE \citep{Hotan04}, TEMPO2 \citep{Hobbs06}, PSRSALSA \citep{Weltevrede2016}}

\end{document}